\documentclass[preprint]{aastex}
\usepackage{color}

\shorttitle{Nuclear gamma-ray line emission}
\shortauthors{Benhabiles et al.}

\begin{document}

\title{Deexcitation nuclear gamma-ray line emission from low-energy cosmic rays in the inner Galaxy}


\author{H. Benhabiles-Mezhoud\altaffilmark{1}}
\affil{Centre de Spectrom\'etrie Nucl\'eaire et de Spectrom\'etrie de Masse, CNRS-IN2P3 and Universit\'e Paris-Sud, 91405 Orsay Campus, France}

\and

\author{J. Kiener} 
\affil{Centre de Spectrom\'etrie Nucl\'eaire et de Spectrom\'etrie de Masse, CNRS-IN2P3 and Universit\'e Paris-Sud, 91405 Orsay Campus, France}
\email{Jurgen.Kiener@csnsm.in2p3.fr}

\and

\author {V. Tatischeff} 
\affil{Centre de Spectrom\'etrie Nucl\'eaire et de Spectrom\'etrie de Masse, CNRS-IN2P3 and Universit\'e Paris-Sud, 91405 Orsay Campus, France}

\and

\author{A. W. Strong}
\affil{Max-Planck Institut f\"{u}r extraterrestrische Physik, 85748 Garching, Germany}
\altaffiltext{1}{present address: IGEE, Universit\'e de M'HAMMED BOUGARA de Boumerd\`es, Boulevard de l'ind\'ependance 35000, Algeria}

\begin{abstract}
Recent observations of high ionization rates of molecular hydrogen in diffuse interstellar clouds point to a distinct low-energy cosmic-ray component. Supposing that this component is made of nuclei, two models for the origin of such particles are explored and  low-energy cosmic-ray spectra are calculated which, added to the standard cosmic ray spectra, produce the observed ionization rates. The clearest evidence of the presence of such low-energy nuclei between a few MeV per nucleon and several hundred MeV per nucleon in the interstellar medium would be a detection of nuclear $\gamma$-ray line emission in the range E$_{\gamma}$ $\sim$ 0.1 - 10 MeV, which is strongly produced in their collisions with the interstellar gas and dust. Using a recent $\gamma$-ray cross section compilation for nuclear collisions, $\gamma$-ray line emission spectra are calculated alongside with the high-energy $\gamma$-ray emission due to $\pi^0$ decay, the latter providing normalization of the absolute fluxes by comparison with Fermi-LAT observations of the diffuse emission above E$_{\gamma}$ = 0.1 GeV. Our predicted fluxes of strong nuclear $\gamma$-ray lines from the inner Galaxy are well below the detection sensitivies of INTEGRAL, but a detection, especially of the 4.4-MeV line, seems possible with new-generation $\gamma$-ray telescopes based on available technology. We predict also strong $\gamma$-ray continuum emission in the 1-8 MeV range, which in a large part of our model space for low-energy cosmic rays exceeds considerably estimated instrument sensitivities of future telescopes. 
 
\end{abstract}


\section{Introduction}

The component below about one GeV per nucleon of the interstellar cosmic-ray spectrum is probably the least well known. Direct observations are not conclusive because cosmic rays with these energies are effectively deflected by the action of the outstreaming solar wind such that the energy spectra measured inside the heliosphere are strongly altered with respect to the interstellar one. It results in an uncertainty of the particle flux which is increasing towards lower energies and below a few hundred MeV per nucleon the cosmic-ray composition and spectrum are essentially unknown. Indirect evidence for an important low-energy component comes from observations of the interstellar ionization rate. In particular, observations of the molecule H$_3^+$ in diffuse interstellar clouds point to a cosmic-ray induced ionization rate which exceeds by more than an order of magnitude the one which is calculated for the standard cosmic-ray flux thought to be produced by diffuse shock acceleration in supernova remnants  (\cite{ind09,ind12}). However, neither the nature - electrons or nuclei - nor the spectrum of the low-energy component can be deduced from these observations. An independent argument for the existence of a significant component of low-energy cosmic-ray nuclei comes from the observed quasi-linear increase of Be with metallicity (\cite{tat11}).

The most compelling indirect observation of this component would be a detection of nuclear $\gamma$-ray lines from the galactic disk. In fact, the strongest lines from the most abundant nuclei are preferentially produced in the cosmic-ray energy range below a few hundred MeV in collisions of protons and $\alpha$-particles with interstellar matter. They are expected to be the same as frequently observed from strong solar flares, i.e. lines from the deexcitation of the first few levels in $^{12}$C, $^{16}$O, $^{20}$Ne, $^{24}$Mg, $^{28}$Si and $^{56}$Fe (\cite{ram79}). \cite{ind09} constructed cosmic-ray spectra which produce the observed ionization rate in diffuse clouds by adding a low-energy component to the standard cosmic-ray spectrum. With that they estimated fluxes of the 4.4-MeV line of $^{12}$C and the 6.1-MeV line of $^{16}$O from the central radian of the Galaxy  close to the sensitivity limits of INTEGRAL. Besides strong narrow lines the nuclear line emission is composed of broad lines produced by the cosmic-ray heavy-ion component and of thousands of weaker lines which together form a quasi-continuum in the E$_\gamma$ $\sim$ 0.1 - 10 MeV range. 
 
In this paper, we calculate this total nuclear $\gamma$-ray line emission from the inner Galaxy by making use of the work of \cite{ind09}, new developments in nuclear reaction cross sections (\cite{mur09}, \cite{ben11}) and recent observations of cosmic-ray induced high-energy $\gamma$-ray emission with Fermi-LAT (\cite{abd09, str11, ack12}). Finally, the possibilities for a detection of this emission with existing and eventual future space-borne $\gamma$-ray instruments are discussed.

\section{Galactic cosmic-ray spectra and composition}

Following the method of \cite{ind09}, the interstellar cosmic-ray (CR) spectra for the calculation of the nuclear $\gamma$-ray line emission have been divided into two distinct components: 

(1) Standard CRs whose origin and acceleration are widely attributed to massive stars through the action of their stellar winds followed by their explosions as supernova. Their composition and energy spectra are relatively well known above several GeV per nucleon from direct measurements while the extrapolation to lower energies can be done with propagation and demodulation calculations. An important indirect constraint on the CR spectra above several hundred MeV per nucleon is the diffuse interstellar high-energy $\gamma$-ray emission due to $\pi^0$ production and decay. This emission has been observed  from several Galactic sources by Fermi-LAT, as for example from the local gas and dust (\cite{abd09}) and from the inner Galaxy (\cite{str11}) which are of particular interest for the present study. These standard CRs produce an  ionization rate well below the mean ionization rate deduced from H$_3^+$ column densities observed in diffuse molecular clouds. 

(2) A distinct low-energy component added to the standard CRs to reach the observed ionization rate. For this we adopt the values deduced from recent observations of the molecule H$_3^+$ in diffuse interstellar clouds pointing to a mean ionization rate of molecular hydrogen of $\mathit \zeta_2$ = 3-4$\times$10$^{-16}$ s$^{-1}$ (\cite{mcc03, ind09, ind12}). Another requirement to this component is that the total interstellar CR spectra stay compatible with the observed CR spectra above 1 GeV per nucleon ensuring also that the high-energy $\gamma$-ray emissions stay compatible with the above cited Fermi-LAT observations. 

\subsection{Standard CRs}

In the most important energy band for comparison with Fermi-LAT data, between about 1 GeV and a few hundred GeV per nucleon, data from some ten  balloon-borne and satellite-borne instruments exist. We adopted the most recent observations of CR fluxes from the PAMELA detector onboard the Russian Resurs-DK1 spacecraft which provides high-precision data for protons and helium nuclei in the rigidity range R $\sim$ 1-1000 GV (\cite{adr11}). Their proton fluxes agree in the kinetic energy range above E $\sim$ 30~GeV, where solar modulation effects should be negligible, with other recent data typically at the 20\% level or better. The agreement is slightly worse for the helium fluxes, where differences up to 40\% can be observed (see e.g. Fig. 1 in \cite{adr11}).     

To obtain the local interstellar (LIS) proton and helium fluxes the observed fluxes have been demodulated with the force-field model of \cite{gle68} with a solar modulation parameter of {$\mathit \Phi$} = 450 MV (\cite{adr11}). This yields LIS proton fluxes in the range {\it E} = 0.89-1010 GeV and LIS helium fluxes in the range {\it E} = 0.36-505 GeV nucleon$^{-1}$. For the LIS helium fluxes we supposed a pure $\alpha$-particle composition, ignoring an eventual small contribution of $^3$He nuclei. 

\begin{figure}
\epsscale{1.0}
\plottwo{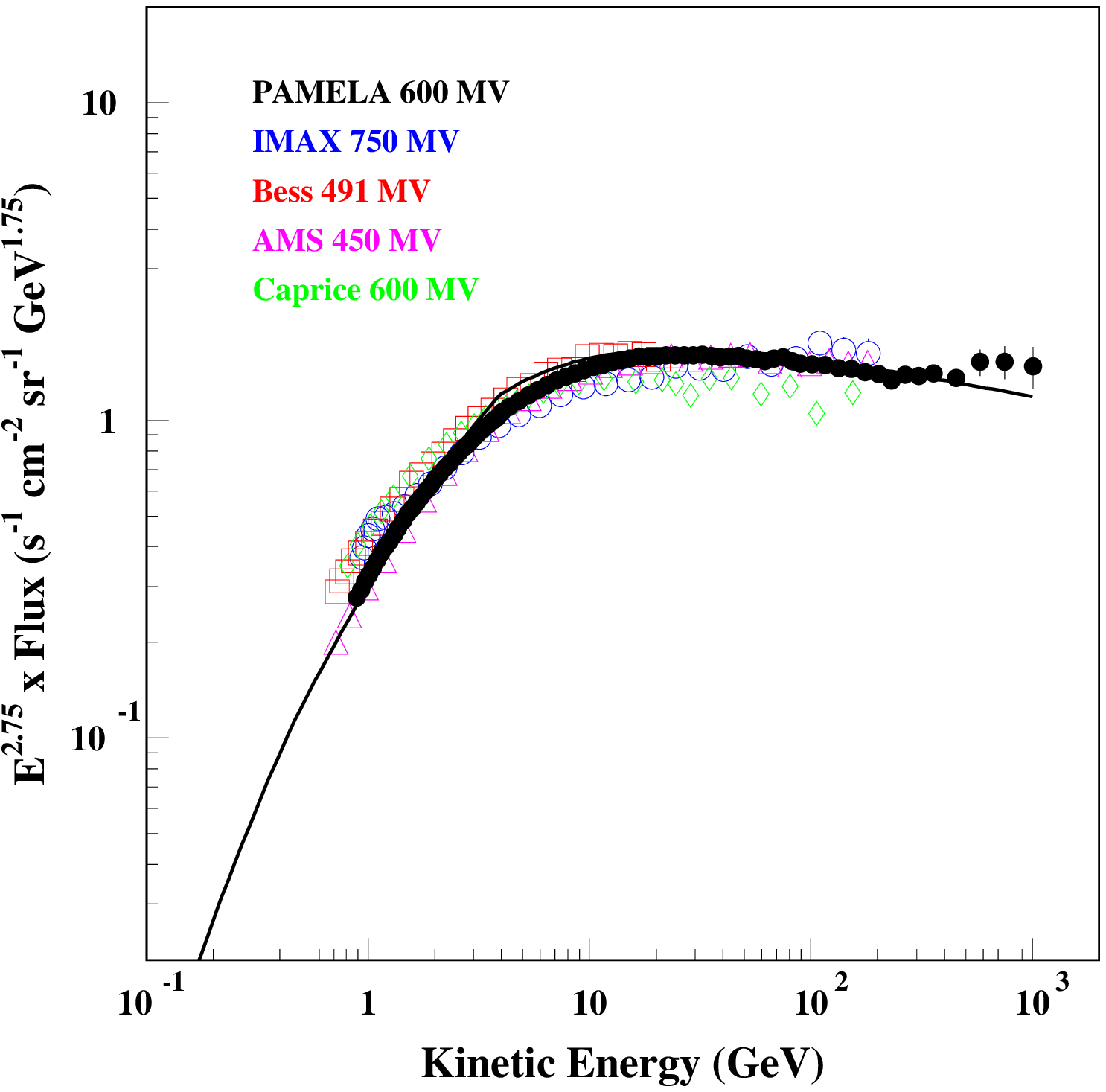}{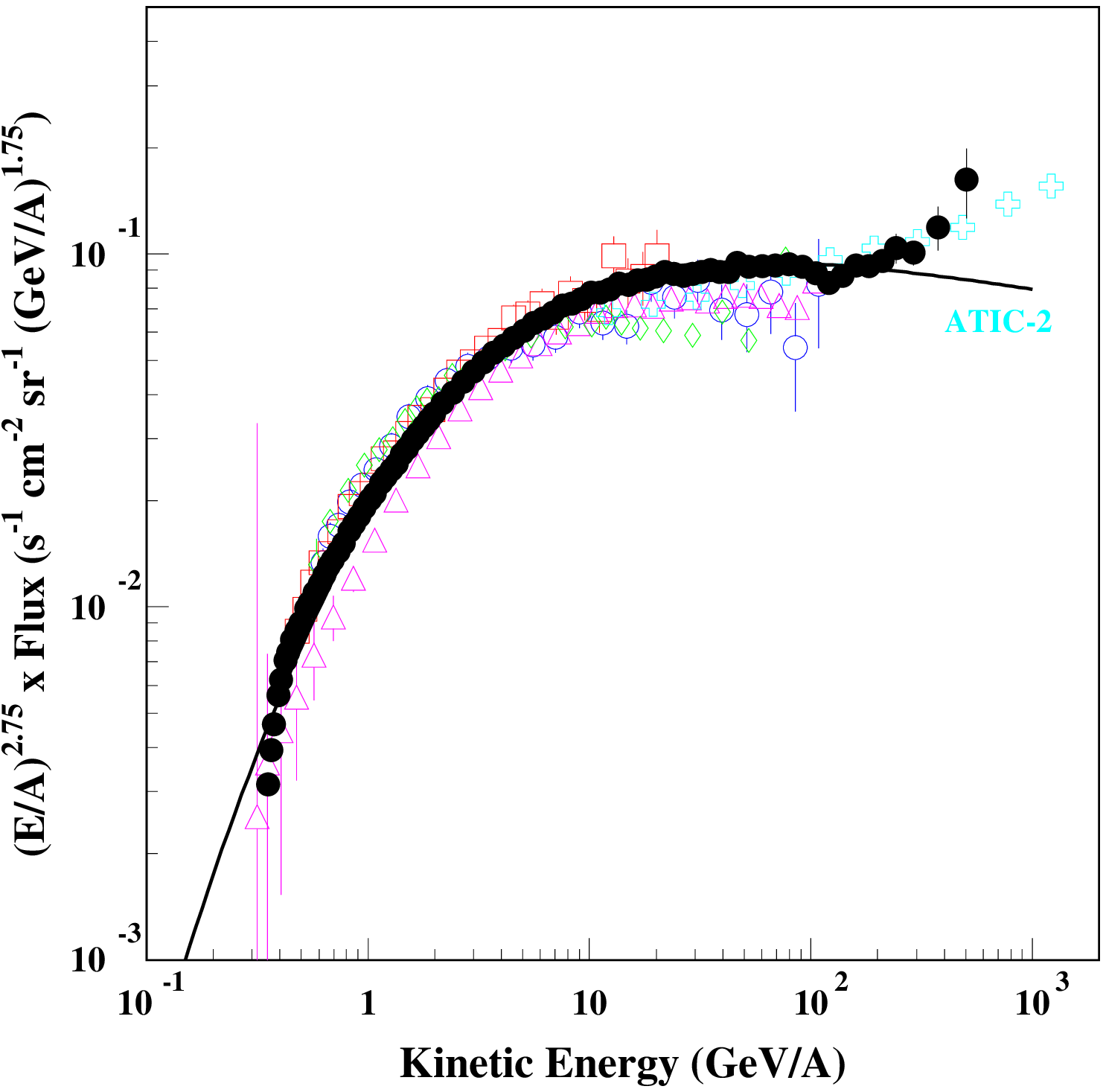}
\caption{Left: LIS proton fluxes from demodulated data of five recent experiments and propagated proton spectrum from the leaky-box model. The modulation potential for the force-field approximation used to demodulate the different data sets are indicated in the left figure. Right: LIS helium fluxes from the same experiments and from ATIC-2. No demodulation was applied to the ATIC-2 data because they start above 10 GeV per nucleon, where modulation should be negligible. \label{hecr}}
\end{figure}

The extrapolation of the proton fluxes to lower energies was done with a standard leaky-box model for Galactic CR propagation. An unbroken power law in particle momentum as it results from diffuse shock acceleration (\cite{je91,der12}) has been used for the source spectrum: {\it F(p)} $\propto$ {\it p}$^{-s}$. Using {\it dF/dE} = {\it dF/dp~ dp/dE} = {\it dF/dp~1/v}, the corresponding energy spectrum is {\it F(E)} = {\it F$_0$ p}$^{-s}$/$\beta$, where {\it v, p} mean the particle velocity and momentum and $\beta$ = {\it v/c}. {\it F$_0$} and the spectral index {\it s} were then adjusted such that the propagated spectrum reproduces the LIS proton fluxes obtained from the PAMELA measurements. In the propagation calculation the escape length distribution was taken from the disk-halo diffusion model of \cite{jon01} and the proton inelastic cross sections from \cite{mos02}. The best adjustment was achieved with a source index {\it s} = 2.35. Figure \ref{hecr} (left panel) shows the propagated proton spectrum together with the Pamela data and some other recent measurements (\cite{men00} [IMAX]; \cite{shi07} [Bess]; \cite{alc0a,alc0b} [AMS]; \cite{boe99} [Caprice]).

For the LIS helium spectrum we took the demodulated helium data of PAMELA  from {\it E} = 0.36 - 300 GeV nucleon$^{-1}$ only, the two data points above that energy suffering a large uncertainty. Extrapolation of the LIS helium spectra to lower energies was done as in the proton case with the leaky-box model with the same parameters for the escape lengths. The inelastic cross section for $\alpha$ + $^4$He was taken from the model of \cite{tri99} except below {\it E}$_{\alpha}$ = 15 MeV where a transition to the experimental values around 10 MeV was operated. Here the best adjustment of the Pamela He data in the range {\it E} = 1-100 GeV nucleon$^{-1}$ was found with a source index {\it s} = 2.32. As shown in Fig. \ref{hecr}, the propagated helium spectrum misses the relatively steep fall of the LIS spectrum below about 0.4 GeV per nucleon. This part is certainly subject to large demodulation uncertainties, the force-field model being  only an approximation and may especially fail at these low energies. In the case of electrons, \cite{str11b} showed recently through constraints on the interstellar electron spectrum from synchrotron radiation that modulation may be significantly overestimated for them and that their true interstellar spectrum shows a sharp turn over below $\sim$1 GeV. However, as this part has only a minor influence on high-energy $\gamma$-ray emission calculations for standard CRs we simply opted for the  values of the propagated spectrum. Above 300 GeV per nucleon we used the data of the ATIC-2 experiment (\cite{pan09}) which have smaller uncertainties than the Pamela data there and agree with them at lower energies. The spectra of heavier nuclei were taken identical to the helium spectrum with relative CR abundances given in table 9.1 of \cite{lon92}, averaged over the three energy bands.

\begin{figure}
\epsscale{1.0}
\plotone{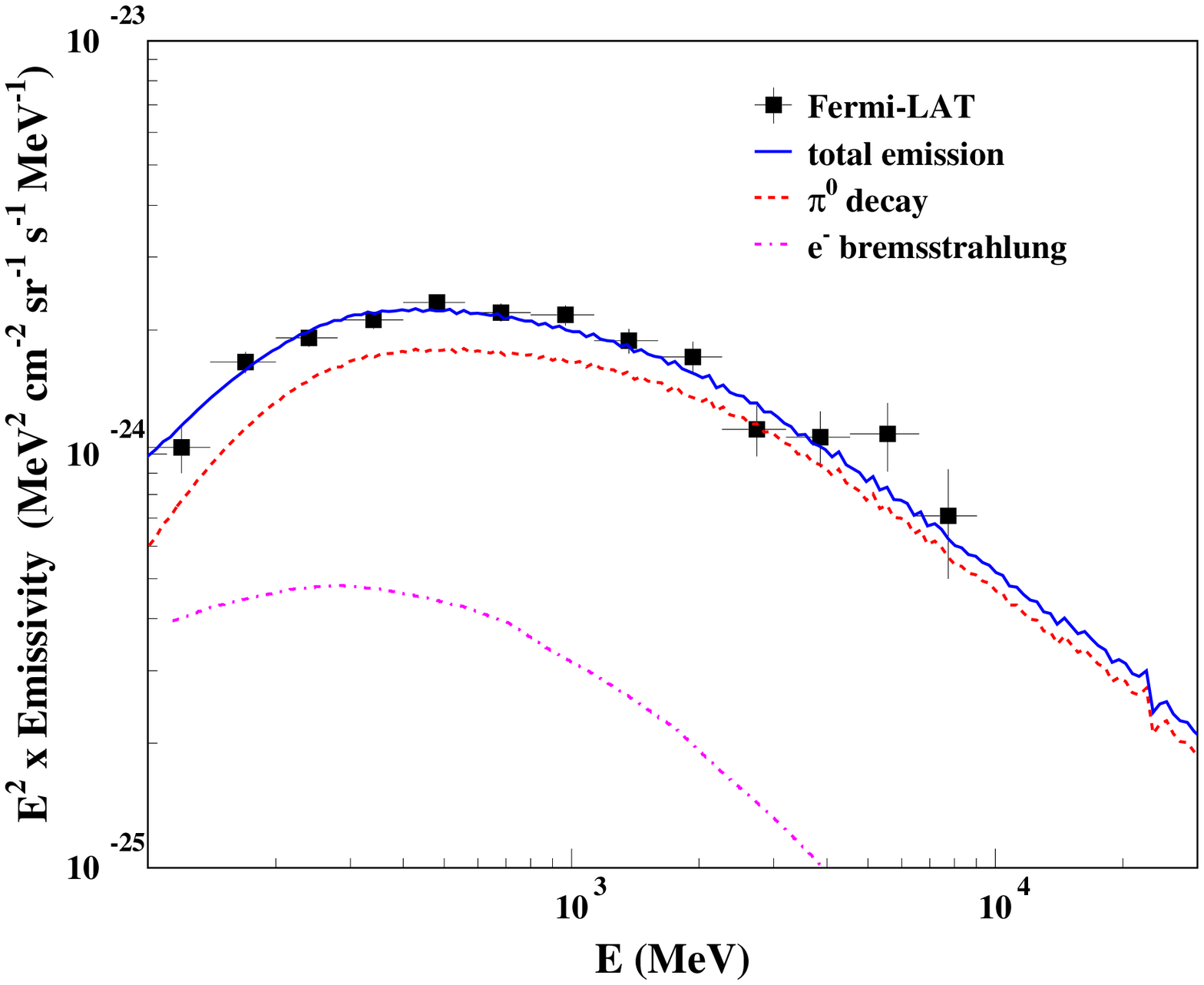}
\caption{$\gamma$-ray emissivity as a function of energy from the local atomic hydrogen gas. Fermi-LAT data and the curve for e$^-$ bremsstrahlung are from \cite{abd09}. The emission from $\pi^0$ decay is calculated with the standard CR composition and spectra as described in Sect. 2.1. \label{lcem} }
\end{figure}

With these CR spectra the $\gamma$-ray emissivity due to $\pi^0$ decay has been calculated with the model of \cite{kam06} for the differential $\gamma$-ray emission cross section in the $\pi^0$ $\rightarrow$ 2$\gamma$ decay. 
Cross section data for $\pi^0$ production in p + p reactions have been taken from the parametrization of \cite{der86}. For the p + $^4$He reaction  the inclusive cross section displayed in Fig. 4 of \cite{mur87} up to 2 GeV has been used. Above 6.3 GeV the cross section has been obtained by multiplying the p + p cross section with the multiplication factor m$_{He p}$ of \cite{mor09} and a linear interpolation was applied in the energy gap. All other reaction channels ($\alpha$ + $^4$He and p,$\alpha$ + HI) were calculated with Mori's multiplication factors relative to the p + $^4$He cross section.  Here, HI stands for elements C, N, O, Ne, Mg, Al, Si, S, Ar, Ca, Fe, and Ni. For the ambient local interstellar composition, solar photospheric abundances have been used. In the CR composition we added $^{11}$B and $^{52}$Cr and $^{55}$Mn as representatives of Sc-Mn, which are relatively abundant in cosmic rays. The result of the calculation is shown in Fig. \ref{lcem} together with the Fermi-LAT data for the local diffuse $\gamma$-ray emissivity of \cite{abd09} and their estimate of the bremsstrahlung component. The calculated curve reproduces remarkably well the observed emissivity in shape and absolute normalization, very close to the calculation of \cite{abd09} albeit with slightly different CR spectra and composition. 

The Galactic CR electron spectrum is constrained by its electromagnetic emissions induced during propagation in the Galaxy and by direct data from many balloon and satellite experiments (see e.g. \cite{str09}). The emissions span a wide range from the Radio band for the synchroton radiation to MeV and TeV energies for bremsstrahlung and inverse Compton scattering. For the standard CR electron spectrum we rely on results of the GALPROP model, which has been developed to calculate numerically the Galactic propagation of hadrons and leptons including secondary-particle production and propagation (\cite{GALPROP}). The model makes also detailed calculations of the electromagnetic emissions which are induced by primary and secondary particles. From the different CR electron spectra and propagation models studied in \cite{str00}, we adopt the conventional model. It fits the direct data above several GeV and agrees simultaneously with the Galactic synchrotron emission. The latter constraint requires a spectral break of the CR electron spectrum below several GeV, which was confirmed in a recent study by \cite{str11b}.   

\subsection{Low-energy part}

The above described standard hadronic CRs yield an ionization rate for molecular hydrogen of  $\mathit \zeta_2$ = 4.3$\times$10$^{-17}$ s$^{-1}$ integrated in the CR energy range {\it E} = 0.002 - 10$^3$ GeV nucleon$^{-1}$, about a factor of ten below the most recent observed mean rates in diffuse molecular clouds $\mathit \zeta_2$ = 4$\times$10$^{-16}$ s$^{-1}$ (\cite{ind09}) and $\mathit \zeta_2$ = 3.5$^{+5.3}_{-3.0}$$\times$10$^{-16}$ s$^{-1}$ (\cite{ind12}). For a direct comparison with \cite{ind09} who first proposed a distinct low-energy CR (LECR) component above 2 MeV per nucleon to account for a high $\mathit \zeta_2$, we adopted in the following for calculations and figures the first value. 

Using the conventional model electron spectrum of \cite{str00} extrapolated down into the eV range, \cite{pad09} found a negligible contribution of electrons to the ionization rate in molecular clouds ($\mathit \zeta_2$ $\sim$10$^{-19}$ s$^{-1}$). There could still be an important CR electron component below 200 MeV to produce the observed ionization rate. Such a LECR electron component has for example been studied by \cite{str00} and \cite{por08} in order to reproduce the Comptel MeV-excess. But this is out of the scope of this paper and we will concentrate in the following on the hadronic component. Supposing that the extra ionization in these clouds is entirely due to low-energy nuclei we explored a range of spectra and compositions for such a LECR component. 

\cite{sch08a} proposed that a CR population of anomalous cosmic rays (ACRs) accelerated in astrospheres of stars similar to the Sun makes up a significant part of the interstellar CR spectrum below {\it E} = 0.3 GeV, and when adding the contributions of other stars those ACRs may  be the dominant component of LECRs. Such a scenario was discussed by \cite{ind09} for their low-energy power-law spectrum (dubbed ''carrot'' spectrum) with power-law index and normalization adjusted to produce the observed ionization rate. We adopted this model in the first scenario for our low-energy part of the interstellar spectrum. 

The spectra and composition were derived in the following way: LECR source spectra were constructed from the ACR proton spectra at the helio- and astropause of \cite{sch08a}, which were constrained by recent data of the Voyager spacecraft. These spectra are well described by a power-law function with exponential cutoff at {\it E$_{c}$: F(E) = F$_0$  E$^{-s}$  e$^{-E/E_{c}}$}   with a spectral index of {\it s} = 2.4.  We explored ACR spectra for this index and an upper and lower limit of {\it s} = 2.7 and {\it s} = 2.0, respectively. For $\alpha$ particles and heavier nuclei the same index was used and the cutoff energy was taken following the species scaling of \cite{ste00}: {\it E$_c$(A,Q)/E$_c$(p) = (A/Q)$^{-2\gamma/(\gamma+1)}$} with energies expressed in energy per nucleon and {\it A, Q} being the mass number and the charge state, respectively. For sake of simplicity we used $\gamma$ = 1 and considered only singly-charged ions, which represent the majority of ACR ions in the heliosphere, reducing the scaling to {\it E$_c$(A,Z)/E$_c$(p) = A$^{-1}$}.  {\it E$_c$} was varied from 5 MeV to an upper limit, where the particle fluxes are still compatible with the observed CR helium and proton fluxes above 0.5 and 1 GeV per nucleon, respectively. The upper limits were found to be {\it E$_c$} = 600, 1200 and 4800 MeV for spectral indices {\it s} = 2.0, 2.4 and 2.7, respectively.  The ACR composition was taken from \cite{cum02}. 

For the second scenario we used source spectra resulting from shock acceleration (SA-LECRs) as for the standard CRs, but with an energy cutoff {\it E$_c$: F(E) = F$_0$ R$^{-s}$/$\beta$ e$^{-E/E_c}$}. {\it R} is the particle rigidity; spectral index {\it s}  and {\it E$_c$}, expressed in energy per nucleon, were taken the same for all species.  For the source spectral index we used {\it s} = 2.35, which is the index found for the standard CRs as our central value, and {\it s} = 2.0 and {\it s} = 2.7 falling in the typical range of models of diffuse shock-acceleration (see e.g. \cite{je91,sch02}). The energy cutoff  {\it E$_c$}, which in this scenario may be considered arising from a tuncation in space or time in the acceleration process was varied from 5 MeV to an upper limit compatible with the observed CR helium and proton fluxes above 0.5 and 1 GeV per nucleon, respectively. It was found to be {\it E$_c$} = 150 MeV for {\it s} = 2.7 and {\it E$_c$} = 120 MeV for {\it s} = 2.35 and {\it s} = 2.0. For the composition of the source spectra we took the CR source abundances of \cite{lun89} from table 9.1 in \cite{rea99}. 

These spectra were then propagated for both scenarios with the leaky-box model to obtain the interstellar LECR spectra. All propagated LECR spectra were normalized to produce an ionization rate $\mathit \zeta_2$ = 3.6$\times$10$^{-16}$ s$^{-1}$ in diffuse clouds, yielding $\mathit \zeta_2$ = 4$\times$10$^{-16}$ s$^{-1}$ when added to the standard CRs. We calculated the ionization rate as \cite{ind09}, assuming in particular that CRs of kinetic energy below a threshold energy of 2 MeV per nucleon cannot penetrate diffuse clouds and thus do not affect the ionization rate there.

\begin{figure}
\epsscale{1.0}
\plotone{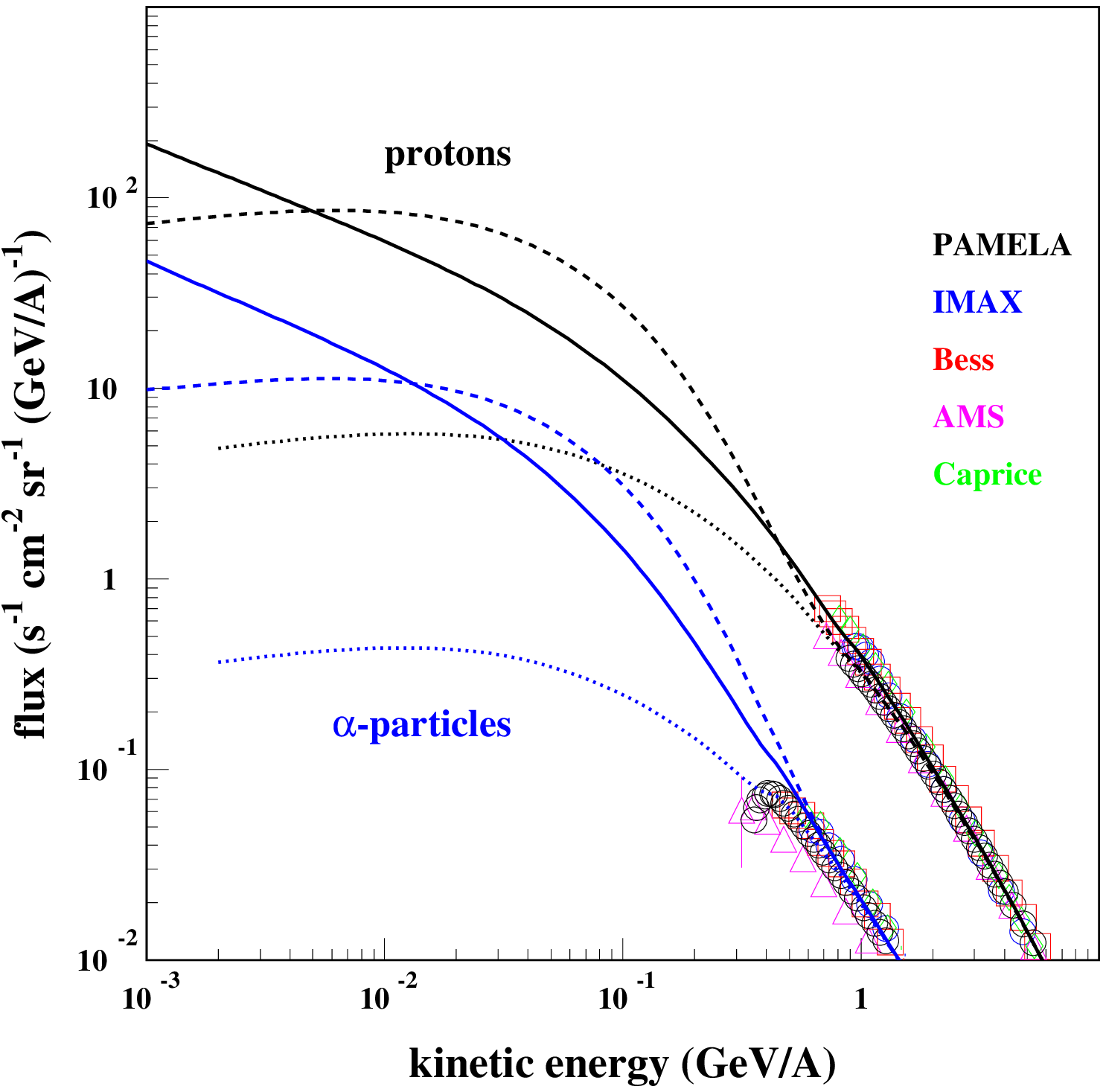}
\caption{Interstellar proton (black lines) and $\alpha$-particle (blue lines) spectra including a low-energy component from propagated source spectra of the model for ACRs of \cite{sch08a} (solid lines) with {\it s} = 2.4, {\it E$_c$} = 1.2 GeV and SA-LECRs (dashed lines) with {\it s} = 2.35,   {\it E$_c$} = 120 MeV. The standard CR spectra from demodulated balloon and satellite data and propagation calculations are shown with the dotted lines. Demodulated data from recent observations are represented by the symbols (see Fig. \ref{hecr} and text for more details). \label{lecr} }
\end{figure}

\begin{figure}
\epsscale{0.8}
\plotone{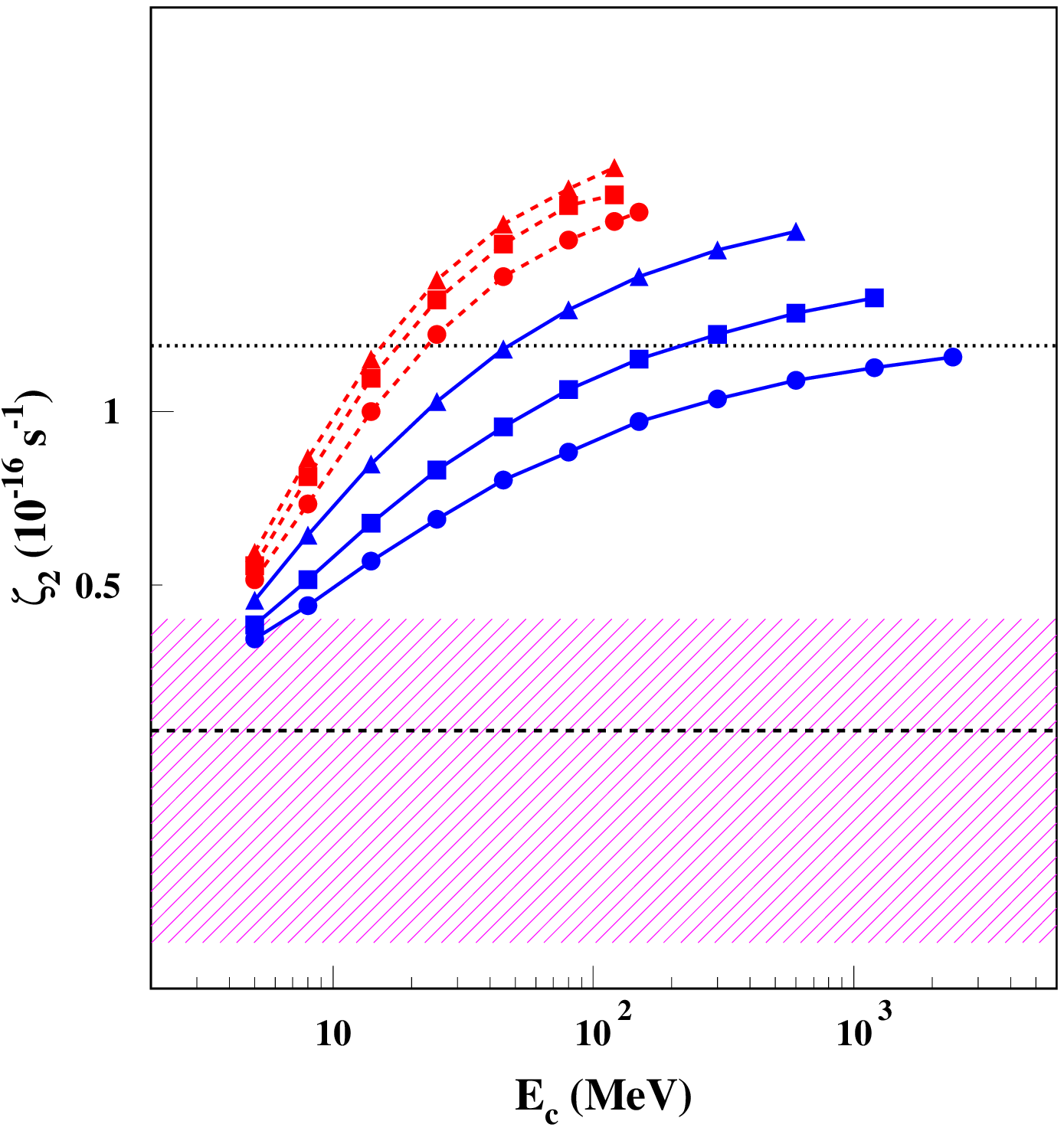}
\caption{Calculated ionization rates of cosmic rays  in dense molecular clouds supposing that particles with energies below 10 MeV per nucleon do not penetrate these places. Red symbols (connected by the dashed lines) show the values for SA-LECRs with spectral indices {\it s} = 2.0 (triangles), {\it s} = 2.35 (squares) and {\it s} = 2.7 (circles), blue symbols (connected by the full lines) the values for ACR-LECRs, {\it s} = 2.0 (triangles), {\it s} = 2.4 (squares) and {\it s} = 2.7 (circles). The ionization rate of standard CRs (0.35$\times$10$^{-16}$ s$^{-1}$) is added. The dashed line and the hatched area show the recommended value of \cite{vdt00} for the cosmic-ray ionization rate and its uncertainty in dense molecular cloud cores ($\mathit \zeta_{CR}$ = (0.28$\pm$0.14) $\times$ 10$^{-16}$s$^{-1}$). The dotted line represents their upper limit ($\sim$1.3$\times$10$^{-16}$ s$^{-1}$). \label{iorates}}
\end{figure}

Examples of LECR proton and $\alpha$-particle spectra are shown in Fig. \ref{lecr} together with the standard CRs alone. The curves with {\it s} = 2.35, {\it E$_c$} = 120 MeV and  {\it s} = 2.4, {\it E$_c$} = 1.2 GeV for SA-LECRs and ACR-LECRs, respectively, represent upper limits still compatible with the demodulated proton and helium data. The ionization rates of LECRs added to the standard CRs in dense clouds, taking a low-energy threshold of 10 MeV as proposed in \cite{ind09}, are displayed in Fig. \ref{iorates}. Nearly all spectra yield rates above the recommended value of \cite{vdt00} for molecular cloud cores and with the exception of ACR-LECRs with {\it s } = 2.7 cross their upper limit for the higher cutoff energies. We note, however, that the energy threshold of 10 MeV is an approximation based on stopping range calculations and certainly subject to considerable uncertainties. Furthermore, magnetic mirroring can decrease the CR ionization rate in molecular cloud cores by a factor of $\sim$3-4 (\cite{pad11}), which would bring all rates if not into the recommended range but at least below the upper limit discussed in \cite{vdt00}.

\section{Nuclear $\gamma$-ray line emission of the inner Galaxy}

Nuclear line emission was calculated for the same reaction channels as for the calculation of $\pi^0$ production except the p + p channel which produces no $\gamma$-ray lines. Cross sections have been taken from  the recent compilation of \cite{mur09}, which contains cross section excitation functions of about 140 relatively strongly produced $\gamma$-ray lines by energetic proton and $\alpha$-particle reactions in astrophysical phenomena like solar flares. The excitation functions extend from reaction threshold to typically 1 GeV per nucleon which in most cases is amply sufficient for the present study where the nuclear $\gamma$-ray line emission is dominated by the low-energy part. Above that range, constant cross sections were supposed. 

For all other $\gamma$-ray lines we relied entirely on calculations with the reaction code TALYS (\cite{talys}) as it was done in \cite{mur09}, with some slight parameter changes for $^{14}$N, $^{20,22}$Ne and $^{28}$Si which gave an improved reproduction of new experimental $\gamma$-ray data for proton and $\alpha$-particle induced reactions with these nuclei (\cite{ben11}). This code works in the range up to 250 MeV for proton reactions and 62.5 MeV per nucleon for $\alpha$-particle reactions. Above these energies a constant cross section was supposed. Lines emitted during the decay of radioactive nuclei were included in these calculations but not the 511-keV line and continuum emissions resulting from $\beta^+$ decay and subsequent positron annihilation. Also emission induced by $\pi^+$-decay positrons was not included in the calculated spectra.

\begin{figure}
\epsscale{1.0}
\plotone{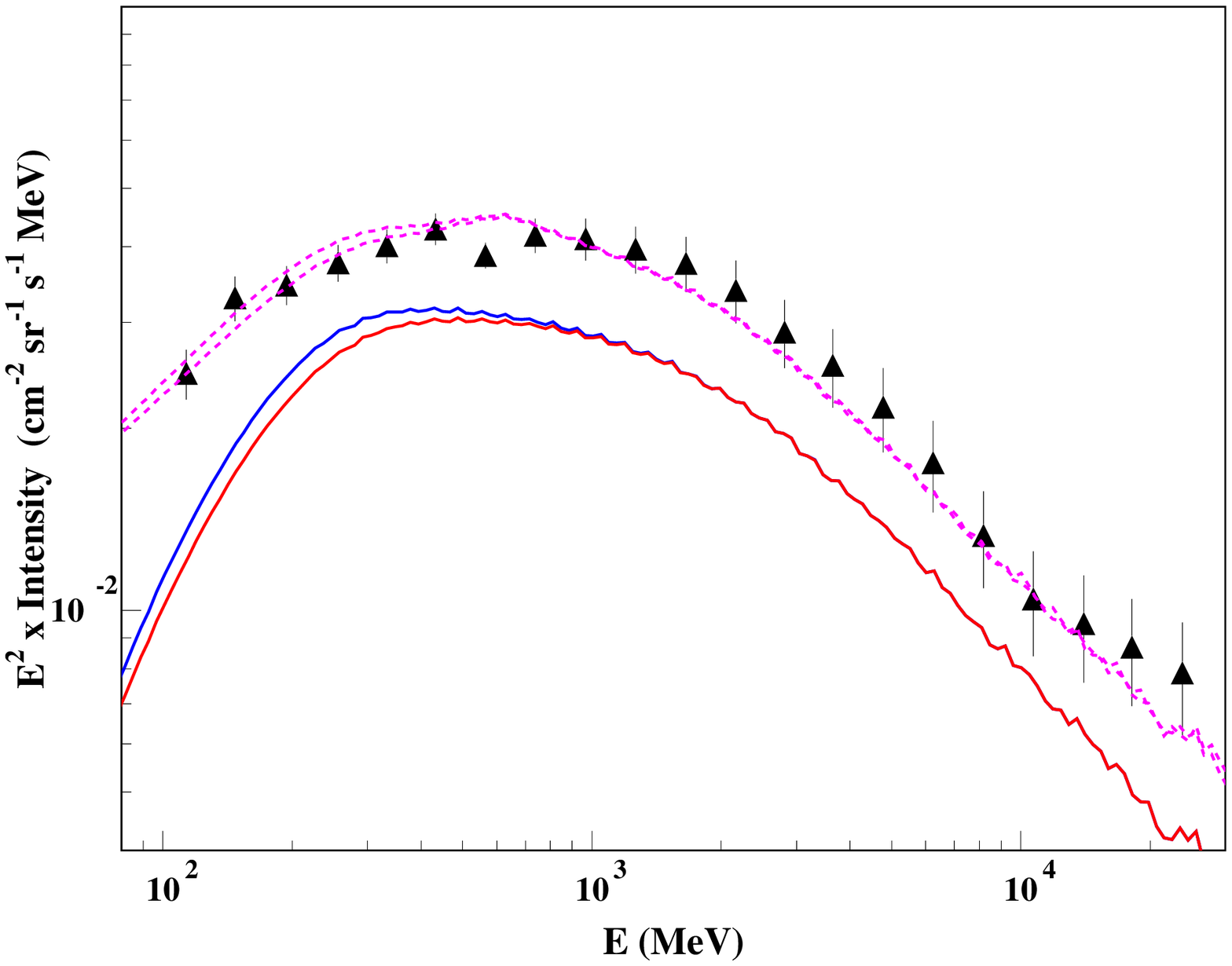}
\caption{Fermi-LAT data of the diffuse $\gamma$-ray emission of the inner Galaxy (300$^{\circ}$ $<$ l $<$ 60$^{\circ}$, $\mid$ b $\mid$ $<$ 10$^{\circ}$) (\cite{str11}) are shown by the black triangles. Solid lines: Calculated $\gamma$-ray emissions from $\pi^0$ decay with CR spectra including a LECR component following the model of \cite{sch08a} for ACRs with {\it s} = 2.4, {\it E$_c$} = 1200 MeV (blue) and {\it s} = 2.4, {\it E$_c$} = 8 MeV (red). The curve for standard CRs is indistinguishable from the red curve, since for {\it E$_c$} = 8 MeV the $\pi^0$ production is entirely due to the standard CRs. The dashed lines show the total emission with contributions of inverse Compton scattering and unresolved point sources as estimated in \cite{str11} added to the calculated $\pi^0$-decay emission. \label{glcfermi}}
\end{figure}

As described above, the cosmic-ray spectra were normalized such that an ionization rate $\mathit \zeta_2$ = 4$\times$10$^{-16}$ s$^{-1}$ was obtained and that the calculated high-energy emission due to $\pi^0$ decay reproduced the Fermi observations of the diffuse $\gamma$-ray emission of the inner Galaxy (\cite{str11}). For the ambient medium composition solar abundances with two times solar metallicity were taken. The Fermi-LAT data and calculations for ACR-LECR spectra with {\it s} = 2.4 and two extreme cases of energy cutoffs {\it E$_c$} = 5 MeV and {\it E$_c$} = 1.2 GeV are shown in Fig. \ref{glcfermi}. When adding the inverse Compton component from high-energy leptons and the estimated emission due to unresolved point sources as presented in \cite{str11} the reproduction is quite good for both cases. The $\pi^0$ production is apparently completely dominated by the standard CRs, even for LECRs with {\it E$_c$} = 1.2~GeV which add only a small component below E $\sim$ 1~GeV.

\begin{figure}
\epsscale{1.0}
\plotone{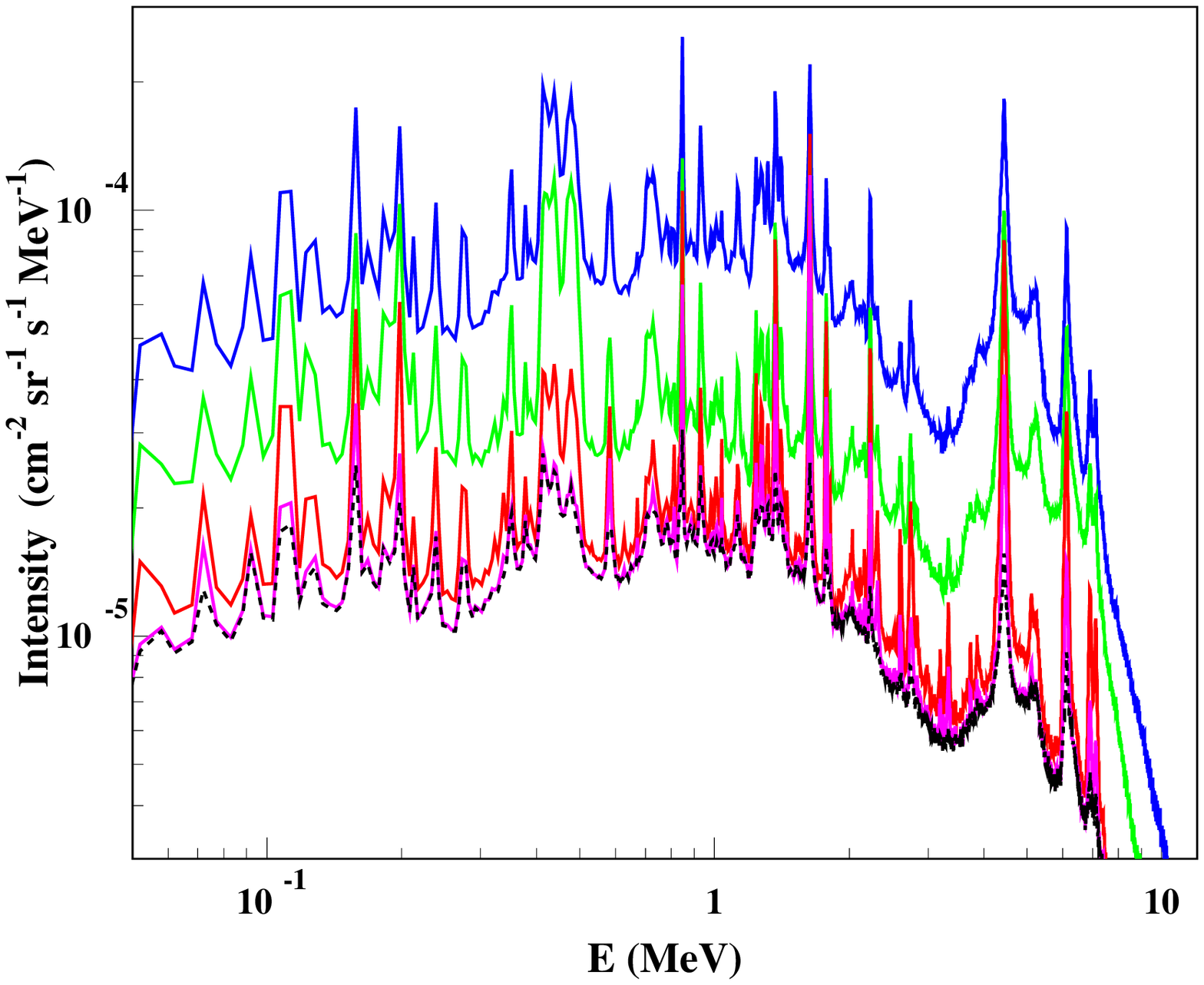}
\caption{Calculated nuclear $\gamma$-ray line emissions from the inner Galaxy for CRs with ACR-LECR components following the model of \cite{sch08a} with {\it s} = 2.4, {\it E$_c$} = 5, 25 and 1200 MeV (magenta, red and green lines, resp.) and SA-LECR with {\it s} = 2.0 and {\it E$_c$} = 120 MeV (blue line). The emission due to the standard CR component alone is shown by the dashed black line. \label{glcnuc}}
\end{figure}

Examples of nuclear $\gamma$-ray line emission for three cases of ACR-LECRs with {\it s} = 2.4 and the case with strongest emission {\it s} = 2.0, {\it E$_c$} = 120 MeV of SA-LECRs are shown in Fig. \ref{glcnuc}. Emission due to the standard CRs alone, as described in Sect. 2.1, is also shown for comparison. The total $\gamma$-ray line intensity for the latter is dominated by interactions of the heavy-ion content due to their enrichment in standard CRs with respect to the interstellar medium. It accounts for about 70\% of the total nuclear line emission. Some outstanding narrow lines, produced in proton- and $\alpha$-particle induced reactions are nevertheless present. The emissions of ACR-LECRs with cutoff energies {\it E$_c$} = 5~MeV and 25 MeV add practically only a narrow-line component, the quasi-continuum being completely dominated by the standard CR component of the CR spectra. In fact, this is mainly because heavy ions play a completely negligible role in low-energy cutoff ACR-LECRs due to their very low cutoff energies ({\it E$_c$(A,Z)/E$_c$(p) = A$^{-1}$}; see Sect. 2.2). Their contribution is less than 10$^{-3}$ at {\it E$_c$} $\le$ 8 MeV, $\sim$6\% at {\it E$_c$} = 80 MeV and raises to $\sim$50\% at {\it E$_c$} = 1.2 GeV.

The overall emission is stronger for SA-LECRs because more particles above the reaction thresholds are present with respect to ACR-LECRs of the same cutoff energy. ACR-LECR spectra with their steep rise towards low energies contain many low-energy particles that contribute strongly to ionization but are very inefficient for $\gamma$-ray line production. This is true for the narrow lines induced by light ions and slightly more pronounced in the quasi-continuum, which reflects furthermore the importance of energetic heavy ions in SA-LECRs. They contribute between $\sim$50\% at {\it E$_c$} = 5 MeV and $\sim$60\% at {\it E$_c$} = 120 MeV to the nuclear $\gamma$-ray emission. 

\begin{figure}
\epsscale{1.0}
\plotone{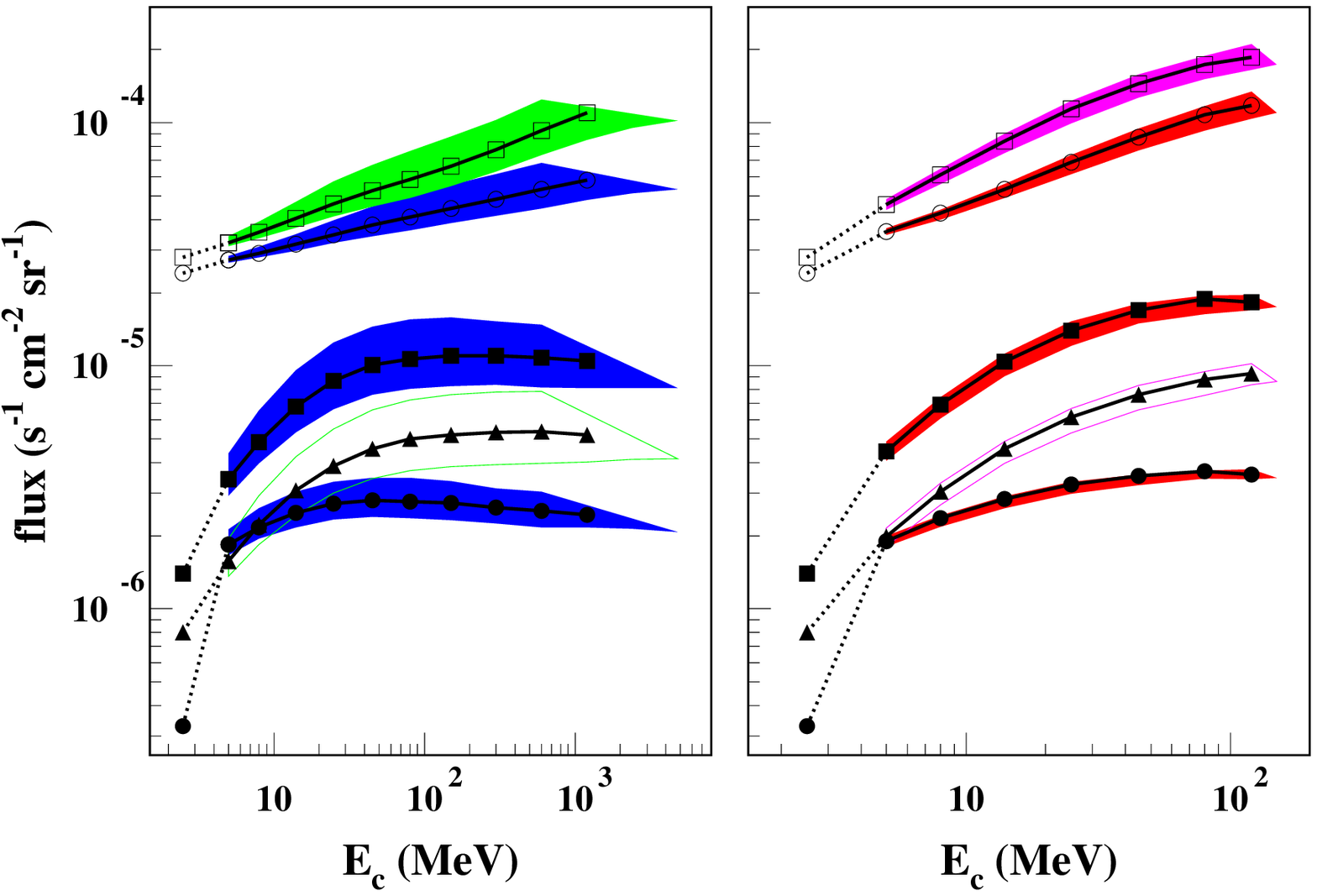}
\caption{Symbols show the fluxes obtained for ACR-LECRs (left figure) with source spectral index of {\it s} = 2.4 and {\it s} = 2.35 for SA-LECRs (right figure), added to the fluxes of the standard CR spectrum: integrated narrow-line fluxes of the 1.63-MeV (filled circles), 6.1-MeV (triangles) and 4.4-MeV lines (filled squares) and integrated fluxes in the 1-3 MeV range (open circles) and 3-8 MeV range (open squares) as a function of the cutoff energy of the LECR component. The filled and open areas show the flux ranges obtained with spectral indices {\it s} = 2.7 (lower flux limit) and {\it s} = 2.0 (higher flux limit) for the LECR source spectra. The values at {\it E$_c$} = 2.5 MeV, connected by dotted lines to the LECR values at 5 MeV correspond to the standard CR component alone. \label{fluxes}}
\end{figure}

The strongest narrow lines in decreasing order are the 4.4-MeV line (essentially from $^{12}$C and $^{11}$B, produced in reactions with $^{12}$C and $^{16}$O), the 6.1-MeV line (mainly inelastic scattering off $^{16}$O and spallation reaction products $^{15}$O and $^{15}$N) and the 1.63-MeV line (mainly from inelastic scattering reactions with $^{20}$Ne). Integrated fluxes of these narrow lines and the integrated fluxes in energy ranges 1-3 MeV and 3-8 MeV are shown on Fig. \ref{fluxes}. Although the continuum fluxes increase steadily with cutoff energy, the narrow-line fluxes tend to level or even slightly decline above {\it E$_c$} $\sim$50~MeV. The latter can be explained by the cross section excitation functions of the dominant line-producing reactions which have their maxima generally well below 50 MeV. An exception are 6.13-MeV fluxes for SA-LECRs which include an important contribution of $^{16}$O spallation reactions with higher thresholds.

Narrow-line and continuum fluxes have also been calculated with two spectra and compositions as proposed by \cite{ind09}. The flux values for their ''carrot'' spectrum are within 20\% of the values for ACR-LECRs with {\it s} = 2.7, {\it E$_c$} = 8 MeV while the values for their power-law spectrum agree to better than 15\% with the ones for SA-LECRs with {\it s} = 2.0, {\it E$_c$} = 14 MeV. The 4.4-MeV and 6.1-MeV narrow line fluxes are smaller by about a factor of ten than the fluxes predicted by \cite{ind09} for the central radian of the Galaxy (330$^{\circ}$ $<$ l $<$ 30$^{\circ}$, $\mid$b$\mid$ $<$ 10$^{\circ}$) probably mainly because of the method of absolute normalization. In \cite{ind09} it was based on the column density of the target material in the central radian assumed to be 10$^{23}$ cm$^{-2}$. By normalizing the nuclear line emission to the $\pi^0$-decay flux from the inner Galaxy as determined by Fermi-LAT (\cite{str11}) our 4.4-MeV and 6.1-MeV line fluxes are independent of the column density. They depend only on the spectral shape and  composition of CRs and the metallicity of the inner Galaxy. Supposing that the intensity of the CRs which are responsible for the bulk of the $\pi^0$ production ({\it E} $\sim$0.5 - 20 GeV per nucleon) is the same throughout the Galaxy, the Fermi-LAT data imply a mean column density of about 2$\times$10$^{22}$ cm$^{-2}$ in the inner Galaxy (300$^{\circ}$ $<$ l $<$ 60$^{\circ}$, $\mid$b$\mid$ $<$ 10$^{\circ}$). This explains at least a large part of the difference. 

\section{Discussion}

Our estimations of nuclear cross section uncertainties can roughly be divided into two categories. (1) Strong lines whose cross sections excitation functions are explicitely listed in the compilation of  \cite{mur09}. They are at least partly based on experimental data whose uncertainties are of the order of 15\% or less. For the 1.63, 4.4 and 6.1-MeV lines, the bulk of the emission is induced by particles in the energy range where a wealth of experimental data exist, and the nuclear uncertainties on the narrow-line fluxes should be less than 20\%. (2) The weak-line quasi-continuum which is taken from TALYS (\cite{talys}) calculations. The predictions of this code have been extensively compared to experimental $\gamma$-ray cross sections in several studies ( \cite{ben11} \cite{mur09} \cite{tat06}) and a typical agreement of a factor of two or better for individual lines was found. As this quasi-continuum consists of thousands of individual lines, the overall uncertainty on the integrated cross section in the two energy bands 1-3 and 3-8 MeV should be smaller than that. A conservative estimate would be less than 50\%. This weak-line emission component is maximal for the high-energy CR component ($\sim$70\%), while for the LECRs it represents $\sim$50\% at the highest-energy cutoffs and at {\it E$_c$} = 5 MeV drops to $\sim$20\% and $\sim$25\% for ACR-LECRs and SA-LECRs, respectively. It results in nuclear uncertainties of the 1-3 and 3-8 MeV fluxes of less than $\sim$40\%.          

Another uncertainty of the narrow-line fluxes comes from the metallicity of the ambient medium. Measurements generally agree for the average metallicity at the Galactocentric distance of the Sun to be equal or slightly below the solar system value (\cite{daf04}, \cite{lem07}, \cite{ped09}, \cite{rol00}). The radial gradients however, especially towards the Galactic center ($R=0$), appear less well defined. Using the available extracted radial gradients, and assuming they stay constant towards the center, extrapolation of element abundances to the center yield values ranging from [O/H] = 0 (\cite{daf04}) to [Fe/H] = 1 (\cite{ped09}), where [O/H]=log10((O/H)$_{R=0}$/(O/H)$_{\odot}$), and (O/H)$_{\odot}$, the solar system value. The fact that the inner Galaxy largely dominates the $\gamma$-ray emission and positive metallicity gradients, however, make a metallicity higher than solar reasonable in this case.  Our choice of taking [X/H] = 0.3, X = C-Ni for the average metallicity towards the inner Galaxy is only a guess with a conservative estimated uncertainty of a factor of two. This uncertainty affects, however, less the quasi-continuum fluxes which are to a great part due to the CR heavy-ion component. Adding that to the nuclear uncertainties a typical overall uncertainty of a factor of two is expected for the calculated narrow-line fluxes and between 50\% and a factor of two for the continuum fluxes. 

The explored parameter space for LECR spectra and composition spans a wide range in spectral forms and composition, from extremely metal-poor for ACRs with low {\it E$_c$} to the metal-rich propagated CR source abundances for SA-LECRs. All spectra are compatible with the high-energy diffuse emission of the inner Galaxy observed by Fermi-LAT, the ionization rate of diffuse clouds, and recent satellite and balloon data of the CR proton and helium spectra. Estimated ionization rates of dense clouds, however, are for all CR spectra, except for the low cutoff-energy ACR-LECRs somewhat higher than the recommended value. These estimations are very uncertain, however, as they neglect the effects of magnetic fields on the CR propagation. \cite{pad11} recently found that magnetic mirroring always dominates over magnetic focusing, implying a reduction of the CR ionization rate in molecular cloud cores by a factor of $\sim$3-4. 

It would of course be very surprising if the LECR component could be described by a single distribution with one definite spectral shape, index, cutoff energy and composition, and be exclusively composed of nuclei. However, whatever the nature of LECRs, they can probably be approximated by a combination of the spectra and compositions studied here and we believe that the nuclear $\gamma$-ray line emission falls into the range which we have explored and which is presented in Figs. \ref{glcnuc},\ref{fluxes}. It is therefore interesting to compare our predictions with $\gamma$-ray instrument sensitivities. 

Integrating over the solid angle of the inner Galaxy (300$^{\circ}$ $<$ l $<$ 60$^{\circ}$, $\mid$b$\mid$ $<$ 10$^{\circ}$) yields flux values (in cm$^{-2}$ s$^{-1}$) 0.73 times the values presented in Fig. \ref{fluxes}. These values are unfortunately far below the narrow-line sensitivities of SPI/INTEGRAL for diffuse emissions: $\sim$5 $\times$ 10$^{-4}$ cm$^{-2}$ s$^{-1}$ for the 4.4- and 6.1-MeV lines with FWHM widths $\Delta$E $\sim$100 keV and $\sim$2 $\times$ 10$^{-4}$ cm$^{-2}$ s$^{-1}$ for the 1.63-MeV line with $\Delta$E $\sim$20 keV (\cite{tew06}). An important fraction of the 6.1-MeV line may be very narrow, if much of the interstellar oxygen is locked up in dust grains (\cite{tat04}). It would result in a sensitivity gain of a factor of 5 at maximum, but even that is still far from the highest estimated fluxes. In summary, no constraints on LECRs can be inferred from  SPI/INTEGRAL observations of the inner Galaxy.

A definite detection of CR-induced nuclear $\gamma$-ray line emission or stringent constraints on the nature of LECRs in the case of non-detection has probably to wait for the next-generation telescopes with an expected increase in sensitivity of typically a factor of 10-100. We therefore compare the calculated fluxes to instrument proposals to the last call (2010) of ESA's 2015-2025 Cosmic Vision program. Of the three instruments with similar estimated sensitivities in the 0.1-10 MeV range which have been presented we focus on the CAPSiTT proposal (\cite{leb10}) for which some of us participated substantially in the performance estimates. 

The most promising narrow line comparing CAPSiTT sensitivities with the predicted fluxes is the 4.4-MeV line, the sensitivity gain for an eventual very-narrow-line component of the 6.1-MeV line is less significant for this instrument with an estimated energy resolution at 6 MeV of $\sim$30 keV. Assuming a uniform emission in the defined region, the CAPSiTT 3$\sigma$ 5-y survey line sensitivity for the 4.4-MeV line ($\Delta$E = 100 keV FWHM) is $\sim$7 $\times$ 10$^{-6}$ cm$^{-2}$ s$^{-1}$ for such an extended emission, using only Compton events in the CAPSiTT instrument. This is at the limit for ACR-LECR spectra and would allow detection for SA-LECRs with cutoff energies above 15 MeV. Pair-creation events, which have not been considered for the proposal to ESA below {\it E$_{\gamma}$} = 10 MeV can improve significantly the sensitivity for such extended emissions, increasing e.g. the effective detection area threefold at {\it E} = 5 MeV. This would allow a detection for nearly all considered LECR scenarios. 

The broadband 1-3 and 3-8 MeV sensitivities in 5 years amount to $\sim$2 $\times$ 10$^{-5}$ cm$^{-2}$ s$^{-1}$ and $\sim$1 $\times$ 10$^{-5}$ cm$^{-2}$ s$^{-1}$, respectively. The flux expected from the standard CRs alone  is already slightly higher than the sensitivity in the 1-3 MeV band and well above the sensitivity in the 3-8 MeV band. A detection looks therefore very promising for all LECR scenarios. It may, however, be difficult to disentangle from other emissions like inverse Compton scattering of CR electrons and unidentified point sources, which probably dominate the diffuse flux of the inner Galaxy in the 1-10 MeV range. Analysis and discussion of the diffuse interstellar emission as measured by SPI/INTEGRAL  and by COMPTEL/CGRO in this energy range can be found in \cite{bou11}.        

The nuclear $\gamma$-ray line emission has probably a latitude and longitude profile similar to that of the high-energy diffuse emission  observed by Fermi-LAT (\cite{str11}). This emission is approximately constant along 320$^{\circ}$ $<$ l $<$ 40$^{\circ}$ and concentrated on the Galactic plane, where about 60\% of the emission is in a band with  b = $\pm$1.5$^{\circ}$. Differences may arise because of the metallicity dependence of narrow-line fluxes, which is negligible for $\pi^0$-decay emission, essentially produced in reactions with ambient H and He. Differences of ionization rates in diffuse interstellar clouds, sometimes observed for sight lines with small angular separation indicate a non-uniform localized distribution of LECRs (\cite{ind12}). This is not surprising since low-energy particles have shorter ranges than the standard CRs resulting certainly also in an inhomogeneous distribution of LECRs in the inner Galaxy, concentrated around the acceleration sites. This would probably affect only slightly the integrated fluxes since they are tied to the mean ionization rate, but the spatial distribution of the nuclear line emission would then be different from the high-energy emission from $\pi^0$ decay. 

These effects would be favorable for an eventual observation by a Compton-imaging telescope like CAPSiTT, whose estimated angular resolution at 5 MeV (0.8$^{\circ}$) is about a factor of three better than that of COMPTEL/CGRO and SPI/INTEGRAL. The point-source sensitivities of CAPSiTT in 5 years amount to $\sim$2 $\times$ 10$^{-7}$ cm$^{-2}$ s$^{-1}$  and $\sim$3 $\times$ 10$^{-7}$ cm$^{-2}$ s$^{-1}$ for the 4.4-MeV line and the 3-8 MeV band, respectively. Our calculated fluxes of the 4.4-MeV line from the inner Galaxy are typically 10-10$^2$ times higher than that. If e.g. the spatial extension of the inhomogeneities is of the order of the angular resolution, a detection gets very probable. Assuming that the large scatter of observed ionization rates in diffuse clouds (see e.g. \cite{ind12}) reflects the LECR density, the 4.4-MeV flux variations may reach a factor of 10. This would be very favorable for a detection by a futur instrument with the characteristics of CAPSiTT. Our calculated fluxes in the 3-8 MeV band are a factor of 10$^2$-10$^3$ higher than the point-source CAPSiTT sensitivity, and a concentration of a few percent of this flux in a point source would be detectable. Following the scatter of ionization rates, a factor-of-few variations in the fluxes are probable although a non-negligible part of the 3-8 MeV band is produced by standard CRs. That inevitably would be detected by a futur instrument with good angular resolution.

\section{Conclusion}

The presented nuclear line and continuum fluxes from the inner Galaxy induced by interactions of CRs in the interstellar medium were calculated for a large variety of LECR spectra and composition and it is probable that the real values fall somewhere inside the obtained ranges. The total nuclear $\gamma$-ray emission from the inner Galaxy is predicted to be in the range (0.1 - 2.0) $\times$ 10$^{-5}$ cm$^{-2}$ s$^{-1}$ for the 4.4-MeV line, (0.1 - 1.0)  $\times$ 10$^{-5}$ cm$^{-2}$ s$^{-1}$ for the 6.1-MeV line and (0.3 - 3.7)  $\times$ 10$^{-6}$ cm$^{-2}$ s$^{-1}$ for the 1.63-MeV line. For the broad-band emissions we find (0.2 - 1.3) and (0.3 - 2.1) $\times$ 10$^{-4}$ cm$^{-2}$ s$^{-1}$ for the 1-3 and 3-8 MeV bands, respectively. An eventual observation of narrow nuclear $\gamma$-ray lines with future $\gamma$-ray space telescopes based on actually available technology appears to be in reach for a large range of LECR scenarios, although challenging for scenarios with low energy cutoffs, where fluxes do not exceed a few 10$^{-6}$ cm$^{-2}$ s$^{-1}$ from the inner Galaxy. It may also be promising for the continuum, especially the 3-8 MeV continuum flux which exceeds the estimated future instrument sensitivities by up to a factor of 20. This is especially true for an instrument with good angular resolution which allows the separation of this emission from other diffuse emissions and a very inhomogeneous spatial distribution of LECRs in the inner Galaxy could not be missed by such an instrument.

\end{document}